\documentclass[aps,prl,twocolumn,showpacs]{revtex4}

\usepackage{epsf}

\begin{document}

\title{Strong exciton-plasmon coupling in semiconducting carbon nanotubes}

\author{I.V.~Bondarev}\email[Corresponding author.
E-mail: ]{ibondarev@nccu.edu}\affiliation{Physics Department,
North Carolina Central University, 1801 Fayetteville Str, Durham,
NC 27707, USA}
\author{K.~Tatur}
\author{L.M.~Woods}\affiliation{Physics Department, University of South Florida,
4202 E.Fowler Ave, Tampa, FL 33620, USA}

\begin{abstract}
We study theoretically the interactions of excitonic states with
surface electromagnetic modes of small-diameter
($\,\lesssim\!1$~nm) semiconducting single-walled carbon
nanotubes. We show that these interactions can result in strong
exciton-surface-plasmon coupling. The exciton absorption
lineshapes exhibit the line (Rabi) splitting
$\sim\!0.1\!-\!0.3$~eV as the exciton energy is tuned to the
nearest interband surface plasmon resonance of the nanotube.~We
expect this effect to open a path to new optoelectronic device
applications of semiconducting carbon nanotubes.
\end{abstract}

\maketitle


Single-walled carbon nanotubes (CNs) are quasi-one-dimensional
(1D) cylindrical wires consisting of graphene sheets rolled-up
into cylinders with diameters $\sim\!1-10$~nm and lengths
$\sim\!1-10^4\,~\mu$m~\cite{Dresselhaus,Dai,Zheng}.~CNs are shown
to~be useful for miniaturized electronic, electromechanical,
chemical and scanning probe devices and as materials for
macroscopic composites~\cite{Baughman}.~The area of their
potential applications has been recently expanded towards
nanophotonics and optoelectronics~\cite{Bondarev06,Bondarev07}
after the experimental demonstration of controllable single-atom
incapsulation into single-walled CNs~\cite{Jeong}.

For pristine (undoped) single-walled CNs, the numerical
calculations predicting large exciton binding energies
($\sim\!0.3\!-\!0.6$~eV) in semiconducting CNs~\cite{Pedersen03}
and even in some small-diameter ($\sim\!0.5$~nm) metallic
CNs~\cite{Spataru04}, followed by the results of various
measurements of the excitonic
photoluminescence~\cite{Wang04,Hagen05,Plentz05}, have become
available. These works, together with other reports investigating
the role of effects such as intrinsic defects~\cite{Hagen05},
exciton-phonon
interactions~\cite{Plentz05,Perebeinos05,Lazzeri05}, external
magnetic and electric fields~\cite{Zaric,Perebeinos07}, reveal the
variety and complexity of the intrinsic optical properties of
carbon nanotubes.

Here we develop a theory for the interactions between excitonic
states and surface electromagnetic modes in small-diameter
($\lesssim\!1$~nm) semiconducting single-walled CNs.~The approach
is based on our recently developed Green function formalism to
quantize an electromagnetic field in the presence of quasi-1D
absorbing bodies~\cite{Bondarev04,Bondarev05,Bondarev06trends}. We
show that such interactions can result in a strong
exciton-surface-plasmon coupling due to the presence of the
low-energy ($\sim\!0.5\!-\!2$~eV) weakly-dispersive interband
plasmon modes~\cite{Pichler98} and the large exciton excitation
energies $\sim\!1$~eV~\cite{Spataru05} in small-diameter CNs.~This
may be used for new CN optoelectronic device applications.

We consider the vacuum-type electromagnetic interaction of an
exciton with surface electromagnetic fluctuations of a
single-walled semiconducting carbon nanotube. No external
electromagnetic field is assumed to be applied. Since the problem
has the cylindrical symmetry, the orthonormal cylindrical basis
$\{\mathbf{e}_{r},\mathbf{e}_{\varphi},\mathbf{e}_{z}\}$ is used
with the vector $\mathbf{e}_{z}$ directed along the nanotube
axis.~The total Hamiltonian of the coupled exciton-photon system
is of the form (we use Gaussian units)
\begin{equation}
\hat{H}=\hat{H}_F+\hat{H}_{ex}+\hat{H}_{int}\,,\label{Htot}
\end{equation}
where the three terms represent the free field, the free exciton,
and their interaction, respectively.~More explicitly, the second
quantized field Hamiltonian
is~\cite{Bondarev04,Bondarev05,Bondarev06trends}
\begin{equation}
\hat{H}_F=\sum_\mathbf{n}\int_0^\infty\!\!\!\!\!d\omega\,\hbar\omega\,
\hat{f}^\dag(\mathbf{n},\omega)\hat{f}(\mathbf{n},\omega)\,,\label{Hf}
\end{equation}
where the scalar bosonic field operators
$\hat{f}^\dag(\mathbf{n},\omega)$ and $\hat{f}(\mathbf{n},\omega)$
create and annihilate, respectively, the surface electromagnetic
excitation of frequency $\omega$ at an arbitrary point
$\mathbf{n}\!=\!\mathbf{R}_n\!=\!\{R_{CN},\varphi_n,z_n\}$
associated with a carbon atom (representing a lattice site) on the
surface of the CN of radius $R_{CN}$. The summation is made over
all the carbon atoms, and in the following it is replaced by the
integration over the entire nanotube surface according to the rule
$\sum_\mathbf{n}\!\ldots\!=\!(1/S_0)\!\int\!d\mathbf{R}_n\!\ldots\!=
\!(1/S_0)\!\int_0^{2\pi}\!d\varphi_nR_{CN}\int_{-\infty}^\infty\!dz_n\!\ldots\,$,
where $S_0\!=\!(3\sqrt{3}/4)b^2$ is the area of an elementary
equilateral triangle selected around each carbon atom in a way to
cover the entire surface of the nanotube with
$b\!=\!1.42$~\AA\space being the carbon-carbon interatomic
distance.

The second quantized Hamiltonian of the free exciton is of the
form (see, e.g., Ref.~\cite{Haken})
\begin{equation}
\hat{H}_F=\sum_{\mathbf{n},f}\!E_f(\mathbf{n})
B^\dag_{\mathbf{n}+\mathbf{m},f}B_{\mathbf{m},f}=
\!\sum_{\mathbf{k},f}\!E_f(\mathbf{k})B^\dag_{\mathbf{k},f}B_{\mathbf{k},f}\,,\label{Hex}
\end{equation}
where the operators $B^\dag_{\mathbf{n},f}$ and $B_{\mathbf{n},f}$
create and annihilate, respectively, an exciton with energy
$E_f(\mathbf{n})$ at the lattice site $\mathbf{n}$ of the CN
surface.~The index $f\,(\ne\!0)$ refers to the internal degrees of
freedom of the exciton.~Alternatively,
$B^\dag_{\mathbf{k},f}\!=\!\sum_{\mathbf{n}}\!B^\dag_{\mathbf{n},f}
e^{i\mathbf{k}\cdot\mathbf{n}}/\sqrt{N}$ creates ($N$ is the
number of the lattice sites) and
$B_{\mathbf{k},f}\!=\!(B^\dag_{\mathbf{k},f})^\dag$ annihilates
the exciton with quasi-momentum
$\mathbf{k}\!=\!\{k_{\varphi},k_{z}\}$ and energy
$E_f(\mathbf{k})\!=\!\sum_\mathbf{n}E_f(\mathbf{n})e^{-i\mathbf{k}\cdot\mathbf{n}}/N
\!=\!E_{exc}^{(f)}+\hbar^2\mathbf{k}^2/2M_{ex}$, where
$E_{exc}^{(f)}\!=\!E_b^{(f)}+E_g$ is the $f$-internal-state
exciton excitation energy, $E_b^{(f)}$ is the corresponding
binding energy of the exciton, $M_{ex}\!=\!m_e+m_h$ is its
effective translational mass, and $E_g$ is the nanotube band
gap.~The two equivalent free-exciton Hamiltonian representations
are related to one another via the obvious orthogonality
relationships
$\sum_\mathbf{n}e^{-i(\mathbf{k}-\mathbf{k}^\prime)\cdot\mathbf{n}}/N\!=
\!\delta_{\mathbf{k}\mathbf{k}^\prime}$ and
$\sum_\mathbf{k}e^{-i(\mathbf{n}-\mathbf{m})\cdot\mathbf{\mathbf{k}}}/N\!=
\!\delta_{\mathbf{n}\mathbf{m}}$ with the $\mathbf{k}$-summation
running over the first Brillouin zone of the nanotube.~The bosonic
field operators in Eq.~(\ref{Hf}) are transformed to the
$\mathbf{k}$-representation in the same way.

The most general (non-relativistic, electric dipole)
exciton-photon interaction on the nanotube surface can be written
in the form (see Refs.~\cite{Bondarev05,Bondarev06trends})
\begin{equation}
\hat{H}_{int}=\hat{H}_{int}^{(1)}+\hat{H}_{int}^{(2)}\label{Hint}
\end{equation}\vspace{-0.5cm}
\[
=-\frac{e}{m_{e}c}\sum_\mathbf{n}\hat{\mathbf{A}}(\mathbf{n})\!\cdot\!
\left[\hat{\mathbf{p}}_\mathbf{n}-\frac{e}{2c}\hat{\mathbf{A}}(\mathbf{n})\right]
+\sum_\mathbf{n}\hat{\mathbf{d}}_\mathbf{n}\!\cdot\mathbf{\nabla}_{\!\mathbf{n}}
\hat{\varphi}(\mathbf{n})\,,
\]
where
$\hat{\mathbf{p}}_\mathbf{n}\!=\!\sum_f\langle0|\hat{\mathbf{p}}_\mathbf{n}|f\rangle
B_{\mathbf{n},f}+h.c.\,$ is the total electron momentum operator
at the lattice site~$\mathbf{n}$ under the optical dipole
transition resulting in the exciton formation at the same site,
$\hat{\mathbf{d}}_\mathbf{n}\!=\!\sum_f\langle0|\hat{\mathbf{d}}_\mathbf{n}|f\rangle
B_{\mathbf{n},f}+h.c.$ is the corresponding dipole moment operator
[related to $\hat{\mathbf{p}}_\mathbf{n}$ via the equation
$\langle0|\hat{\mathbf{p}}_\mathbf{n}|f\rangle\!=\!im_{e}E_{f}(\mathbf{n})
\langle0|\hat{\mathbf{d}}_\mathbf{n}|f\rangle/\hbar e$].~The
vector potential operator $\hat{\mathbf{A}}(\mathbf{n})$ (the
Coulomb gauge is assumed) and the scalar potential operator
$\hat{\varphi}(\mathbf{n})$ represent, respectively, the
nanotube's transversely polarized surface electromagnetic modes
and longitudinally polarized surface electromagnetic modes which
the exciton interacts with. We express them in terms of our
earlier developed electromagnetic field quantization formalism in
the presence of quasi-1D absorbing bodies to
obtain~\cite{Bondarev05,Bondarev06trends}
\begin{equation}
\hat{\mathbf{A}}(\mathbf{n})=\int_0^\infty\!\!\!\!\!d\omega\frac{c}{i\omega}\,
\hat{\underline{\mathbf{E}}}^\perp(\mathbf{n},\omega)+h.c.=
\sum_{\mathbf{m}}\int_0^\infty\!\!\!d\omega\label{A}
\end{equation}\vspace{-0.5cm}
\[
\times\frac{4}{c}\sqrt{\pi\hbar\omega\mbox{Re}\sigma_{zz}(R_{CN},\omega)}\;
^{\perp}G_{zz}(\mathbf{n},\mathbf{m},\omega)\hat{f}(\mathbf{m},\omega)+h.c.
\]
and
\begin{equation}
-\mathbf{\nabla}_{\!\mathbf{n}}\hat{\varphi}(\mathbf{n})=\int_0^\infty\!\!\!\!\!d\omega\,
\hat{\underline{\mathbf{E}}}^\parallel(\mathbf{n},\omega)+h.c.=
\sum_{\mathbf{m}}\int_0^\infty\!\!\!d\omega\label{phi}
\end{equation}\vspace{-0.5cm}
\[
\times\frac{4i\omega}{c^2}\sqrt{\pi\hbar\omega\mbox{Re}\sigma_{zz}(R_{CN},\omega)}\;
^{\parallel}G_{zz}(\mathbf{n},\mathbf{m},\omega)\hat{f}(\mathbf{m},\omega)+h.c.
\]
with the total electric field operator given by
$\hat{\mathbf{E}}(\mathbf{n})\!=\!\int_0^\infty\!\!d\omega
[\hat{\underline{\mathbf{E}}}^\perp(\mathbf{n},\omega)+
\hat{\underline{\mathbf{E}}}^\parallel(\mathbf{n},\omega)]+h.c.$,
$^{\perp(\parallel)}G_{zz}(\mathbf{n},\mathbf{m},\omega)$ being
the zz-component of the transverse (longitudinal) Green tensor
(with respect to the first variable) of the electromagnetic
subsystem, and $\sigma_{zz}(R_{CN},\omega)$ representing the CN
dynamic surface axial conductivity per unit length.

Equations~(\ref{Htot})--(\ref{phi}) form the complete set of
equations describing the exciton-photon coupled system on the CN
surface in terms of the electromagnetic field Green tensor and the
CN surface axial conductivity. The conductivity is found
beforehand from the realistic band structure of a particular CN.
The Green tensor is derived by expanding the solution of the Green
equation in cylindrical coordinates and determining the Wronskian
normalization constant from the appropriately chosen boundary
conditions on the CN surface (see
Refs.~\cite{Jackson,Bondarev04,Bondarev05,Bondarev06trends}).~More
details on these calculations will be given elsewhere.

It is important to realize that the transversely polarized surface
electromagnetic mode contribution to the interaction Hamiltonian
from Eq.~(\ref{Hint}) (first term) is negligible compared to the
longitudinally polarized surface electromagnetic mode contribution
(second term). The point is that, because of the nanotube
quasi-one-dimensionality, the exciton quasi-momentum vector and
all the relevant vectorial matrix elements of the momentum and
dipole moment operators are directed predominantly along the CN
axis (the longitudinal exciton). This prevents the exciton from
the electric dipole coupling to transversely polarized surface
electromagnetic modes as they propagate predominantly along the CN
axis with their electric vectors orthogonal to the propagation
direction. The longitudinally polarized surface electromagnetic
modes are generated by the electronic Coulomb potential (see,
e.g., Ref.~\cite{Landau}), and therefore represent CN surface
plasmon excitations.~These have their electric vectors directed
along the propagation direction. They do couple to the
longitudinal excitons on the CN surface. Such modes were observed
in Ref.~\cite{Pichler98}. They occur in CNs both at high energies
(well-known $\pi$-plasmon at $\sim\!6$~eV) and at comparatively
low energies of $\sim\!0.5\!-2$~eV [plasmon modes associated with
(transversely quantized) interband electronic transitions].

To obtain the dispersion equation of the coupled exciton-plasmon
excitations, we utilize Bogoliubov's ca\-no\-nical transformation
technique (see, e.g., Ref.~\cite{Davydov}) and diagonalize the
Hamiltonian~(\ref{Htot})--(\ref{phi}) exactly. This results in
\begin{equation}
\hat{H}=\sum_{\mathbf{k},\,\mu=1,2}\!\!\!\!\!\hbar\omega_\mu(\mathbf{k})\,
\hat{\xi}^\dag_\mu(\mathbf{k})\hat{\xi}_\mu(\mathbf{k})+E_0\,,
\label{Htotdiag}
\end{equation}
where
$\hat{\xi}_\mu(\mathbf{k})\!=\!\sum_{f}[u_\mu^\ast(\mathbf{k},\omega_f)B_{\mathbf{k},f}
-v_\mu(\mathbf{k},\omega_f)B_{-\mathbf{k},f}^\dag]+
\int_0^\infty\!d\omega[u_\mu(\mathbf{k},\omega)\hat{f}(\mathbf{k},\omega)-
v_\mu^\ast(\mathbf{k},\omega)\hat{f}^\dag(-\mathbf{k},\omega)]$
annihilates and
$\hat{\xi}^\dag_\mu(\mathbf{k})\!=\![\hat{\xi}_\mu(\mathbf{k})]^\dag$
creates the coupled exciton-plasmon excitations. Here $\mu$
enumerates the two exciton-plasmon branches, $u_\mu$ and $v_\mu$
are the (appropriately chosen) canonical transformation
coefficients, $\omega_f\!=\!E_f/\hbar$. The energy $E_0$ is the
"vacuum" energy with no exciton-plasmons excited in the system,
and $\hbar\omega_\mu(\mathbf{k})$ is the exciton-plasmon energy
given by the solution of the following (dimensionless) dispersion
equation

\begin{equation}
x_\mu^2-\varepsilon_f^2-\frac{\varepsilon_f}{2\pi}\int_0^\infty\!\!\!\!\!dx\,
x\,\bar\Gamma_0^f(x)\frac{\rho(x)}{x_\mu^2-x^2}=0\,.\label{dispeq}
\end{equation}
Here $x\!=\!\hbar\omega/2\gamma_0$,
$x_\mu\!=\!\hbar\omega_\mu(\mathbf{k})/2\gamma_0$,
$\varepsilon_f\!=\!E_f(\mathbf{k})/2\gamma_0$ with
$\gamma_0\!=\!2.7$~eV being the carbon nearest neighbor overlap
integral entering the CN surface axial conductivity $\sigma_{zz}$.
The function
$\bar\Gamma_0^f(x)\!=\!(2\gamma_0/\hbar)^2\,(4|d_f|^2x^3/3\hbar
c^3)$, where
$d_f\!=\!\sum_{\mathbf{n}}\langle0|\hat{\mathbf{d}}_\mathbf{n}|f\rangle$,
represents the (dimensionless) spontaneous decay rate, and
\begin{equation}
\rho(x)=\frac{3S_0}{4\pi\alpha
R_{CN}^2}\;\mbox{Re}\frac{1}{\bar\sigma_{zz}(x)}\label{plDOS}
\end{equation}
stands for the surface plasmon density of states (DOS) responsible
for the decay rate variation due to the coupling to plasmon
modes.~Here $\alpha\!=\!e^2/\hbar c\!=\!1/137$ is the
fine-structure constant and
$\bar\sigma_{zz}\!=\!2\pi\hbar\sigma_{zz}/e^2$ is the
dimensionless CN surface axial conductivity per unit length.~The
conductivity factor equals
$\mbox{Re}(1/\sigma_{zz})\!=\!-(8\pi/\omega
R_{CN})\mbox{Im}[1/(\epsilon_{zz}-1)$ in view of the Drude
relationship $\sigma_{zz}\!=\!-i\omega(\epsilon_{zz}-1)/4\pi
S\rho_{T}$, where $\epsilon_{zz}$ is the longitudinal (along the
CN axis) dielectric function, $S$ and $\rho_{T}$ are the surface
area of the tubule and the number of tubules per unit volume,
respectively~\cite{Bondarev04,Bondarev05,Bondarev06trends}. This
relates very closely the surface plasmon DOS
function~(\ref{plDOS}) to the loss function
$-\mbox{Im}(1/\epsilon)$ measured in Electron Energy Loss
Spectroscopy (EELS) experiments to determine the properties of
collective electronic excitations~\cite{Pichler98}.

\begin{figure}[t]
\epsfxsize=8.65cm\centering{\epsfbox{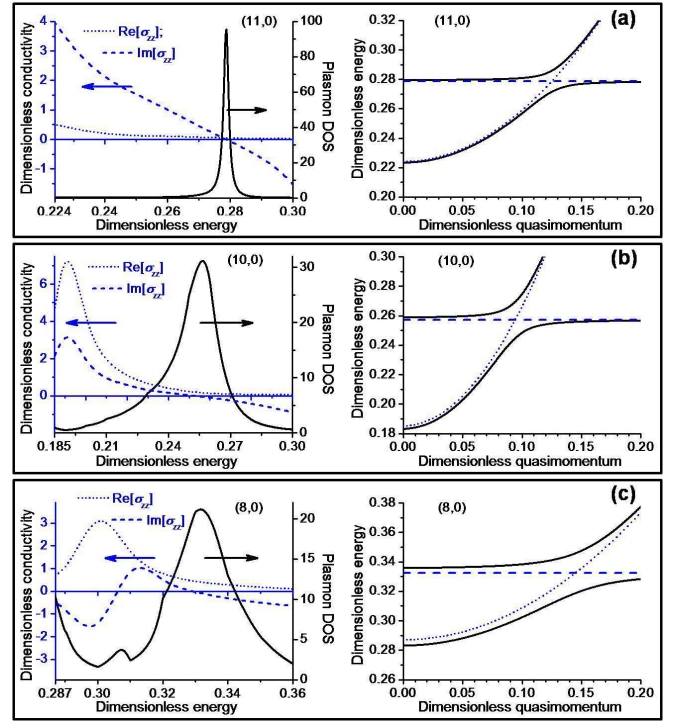}}\vskip-0.2cm
\caption{(Color online)~(a),(b),(c)~Surface plasmon DOS and
conductivities (left panels), and lowest bright exciton dispersion
when coupled to plasmons (right panels) in (11,0), (10,0) and
(8,0) CN, respectively.~Dimensionless energy is defined as
[\emph{Energy}]/$2\gamma_0$; see text for dimensionless
quasimomentum.} \label{fig1}
\end{figure}

The calculated plasmon DOS functions are shown in the left panels
of Fig.~\ref{fig1}~(a),~(b) and~(c), along with the functions
$\mbox{Re}\,\bar\sigma_{zz}(x)$ and
$\mbox{Im}\,\bar\sigma_{zz}(x)$, for the (11,0), (10,0) and (8,0)
CN we study here ($R_{CN}\!=\,$0.43, 0.39 and 0.31~nm,
respectively).~In all graphs the lower dimensionless energy limits
are set up to equal the lowest bright exciton excitation energy
[$E_{exc}\!=\,$1.21, 1.00 and 1.55~eV for (11,0), (10,0) and (8,0)
CN~\cite{Spataru05}, respectively]. Peaks of the plasmon DOS are
seen to coincide in energy with the zeros of the imaginary
conductivities (or the zeros of the real dielectric functions),
clearly indicating the plasmonic nature of the CN surface
excitations under consideration.~Peaks broaden as the CN diameters
decrease.~This is consistent with the stronger hybridization
effects in smaller-diameter CNs~\cite{Blase94}.~The CN dielectric
response is calculated in the following way. First, we adapt the
nearest-neighbor non-orthogonal tight-binding
approach~\cite{Valentin} to determine the realistic band structure
of each CN. Then, the room-temperature longitudinal dielectric
function and conductivity are calculated within the random-phase
approximation with electronic dissipation processes included in
the relaxation-time approximation (electron scattering length of
$130R_{CN}$ was used~\cite{Lazzeri05}).

We further take advantage of the peak structure of $\rho(x)$ and
solve Eq.~(\ref{dispeq}) analytically using the Lorentzian
approximation with $\rho(x)\!\approx\!\rho(x_p)\Delta
x_{p}^2/[(x-x_{p})^2+\Delta x_{p}^2]$, where $x_{p}$ and $\Delta
x_{p}$ are, respectively, the position and the
half-width-at-half-maximum of the plasmon resonance
(Fig.~\ref{fig1}, left panels) closest to the lowest bright
exciton excitation energy in the same nanotube. We obtain
\begin{equation}
x_{1,2}=\sqrt{\frac{\varepsilon_f^2+x_{p}^2}{2}\pm\frac{1}{2}
\sqrt{(\varepsilon_f^2\!-x_{p}^2)^2+F_{\!p}\,\varepsilon_f}}\,,
\label{dispsol}
\end{equation}
$F_{\!p}\!=\!F(x_p)\Delta x_p(1\!-\!\Delta x_{p}/\pi x_{p})$,
$F(x_p)\!=\!2x_p\bar\Gamma_0^f(x_p)\rho(x_p)$. The dispersion
curves thus obtained are shown in the right panels of
Fig.~\ref{fig1}~(a),~(b) and~(c) as functions of the longitudinal
dimensionless quasi-momentum. All graphs demonstrate a clear
anticrossing behavior of the two exciton-plasmon branches with the
(Rabi) energy splitting $\sim\!0.1$~eV. In these calculations, we
estimated the interband transition matrix element in
$\bar\Gamma_0^f(x_p)$ from the relationship
$|d_f|^2\!=\!3\hbar\lambda^3/4\tau_{ex}^{rad}$~\cite{Hanamura},
where $\tau_{ex}^{rad}$ is the exciton intrinsic radiative
lifetime, $\lambda\!=\!2\pi c\hbar/E$ with
$E\!=\!E_{exc}+(2\pi\hbar/3b)^2t^2/2M_{ex}$ being the total energy
and $-1\!\le\!t\!\le\!1$ representing the longitudinal
dimensionless quasi-momentum of the exciton.~We used the lowest
bright exciton parameters reported in Ref.~\cite{Spataru05}.

\begin{figure}[t]
\epsfxsize=8.65cm\centering{\epsfbox{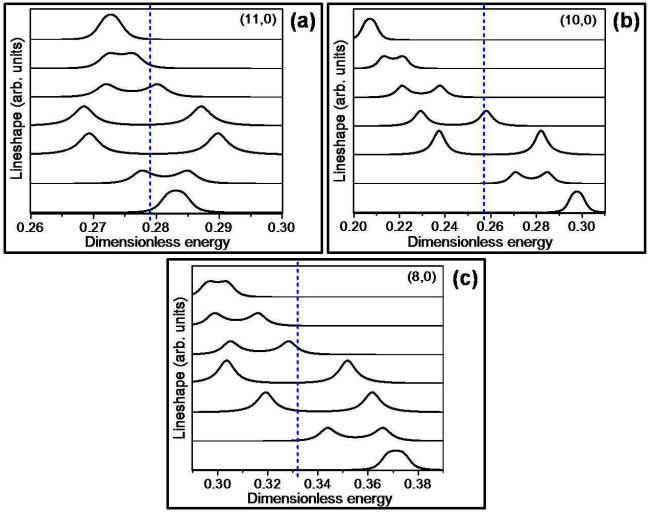}}\vskip-0.2cm
\caption{(Color online)~(a),(b),(c)~Exciton absorption lineshapes
as the exciton energies are tuned to the nearest plasmon resonance
energies (vertical dashed lines here; see Fig.~\ref{fig1}, left
panels) in the (11,0), (10,0) and (8,0) CN,
respectively.}\label{fig2}\vspace{-0.5cm}
\end{figure}

We also derive the (dimensionless) exciton absorption lineshape
function $I(x)$ in the vicinity of the plasmon resonance in the
way this was done in Ref.~\cite{Bondarev06} for the optical
absorption by atomically doped CNs. In doing so, we take into
account the exciton-phonon scattering in the relaxation time
approximation. We obtain

\begin{equation}
I(x)=I_{0}(\varepsilon_f)\frac{(x-\varepsilon_f)^{2}+\Delta_1^2}
{[(x-\varepsilon_f)^{2}-X_f^{2}/4]^{2}+(x-\varepsilon_f)^{2}\Delta_2^2}\,,
\label{Ixfin}
\end{equation}
where
$I_{0}(\varepsilon_f)\!=\!\bar\Gamma_0^f(\varepsilon_f)\rho(\varepsilon_f)/2\pi$,
$X_f\!=\!\sqrt{4\pi\Delta x_p\,I_{0}(\varepsilon_f)}\,$,
$\Delta_1\!=\!\sqrt{2\Delta\varepsilon_f+\Delta x_p}\,$, and
$\Delta_2\!=\!\sqrt{(\Delta\varepsilon_f)^2+\Delta_1^2}$ with
$\Delta\varepsilon_f\!=\!\hbar/2\gamma_0\tau_{ph}$ being the
exciton energy broadening due to the phonon scattering with the
relaxation time $\tau_{ph}$. The calculated exciton absorption
lineshapes for the CNs under consideration are shown in
Fig.~\ref{fig2}~(a),~(b) and~(c) as the exciton energies are tuned
to the nearest plasmon resonances.~We used $\tau_{ph}\!=\!30$~fs
as reported in Ref.~\cite{Perebeinos05}. The clear line (Rabi)
splitting effect is seen to be of the order of $\,0.1-0.3$~eV,
indicating the strong exciton-plasmon coupling with the formation
of the mixed surface plasmon-exciton excitations. The splitting is
larger in the smaller diameter nanotubes, and is not masked by the
exciton-phonon scattering.

In summary, we have shown the strong exciton-surface-plasmon
coupling effect with characteristic Rabi splitting
$\sim\!0.1\!-\!0.3$~eV in small-diameter ($\lesssim\!1$~nm)
semiconducting CNs.~This is almost as large as typical exciton
binding energies in such CNs
($\sim\!0.3\!-\!0.6$~eV~\cite{Pedersen03}), and of the same order
of magnitude as the exciton-plasmon Rabi splitting in organic
semiconductors ($\sim\!180$~meV~\cite{Bellessa}). Also, this is
much larger than the exciton-polariton Rabi splitting in
semiconductor microcavities
($\sim\!140-400\,\mu\mbox{eV}\,$\cite{Reithmaier}), or the
exciton-plasmon Rabi splitting in hybrid semiconductor-metal
nanoparticle molecules~\cite{Govorov}.~However, the formation of
the strongly coupled exciton-plasmon states is only possible if
the exciton total energy is in resonance with the energy of an
interband surface plasmon mode. That is not always the case
experimentally.~To realize the strong coupling effect, the exciton
energy should be tuned to the nearest plasmon resonance in ways
used for~ the excitons in semiconductor microcavities
--- thermally~\cite{Reithmaier} (by elevating sample temperature),
and/or electrostatically~\cite{Zrenner} (via the quantum confined
Stark effect with an external electrostatic field applied
perpendicular to the nanotube axis).~This opens paths to new
optoelectronic device applications of semiconducting CNs.

The work is supported by NSF (grant ECS-0631347). K.T. and L.M.W.
are supported by DOE (grant DE-FG02-06ER46297).

\end{document}